\documentclass[12pt,a4paper]{article}
\usepackage{amsfonts,latexsym}
\usepackage{amsmath,amssymb}
\usepackage{graphicx,color}

\renewcommand{\thefootnote}{\fnsymbol{footnote}}

\begin{document}

\vspace{12mm}

\begin{center}
{{{\Large {\bf  Black hole with primary scalar hair \\ in Einstein-Weyl-Maxwell-conformal scalar theory }}}}\\[10mm]

{De-Cheng Zou$^{a,b}$\footnote{e-mail address: dczou@yzu.edu.cn} and Yun Soo Myung$^a$\footnote{e-mail address: ysmyung@inje.ac.kr}}\\[8mm]

{${}^a$Institute of Basic Sciences and Department  of Computer Simulation, Inje University Gimhae 50834, Korea\\[0pt] }

{${}^b$Center for Gravitation and Cosmology and College of Physical Science and Technology, Yangzhou University, Yangzhou 225009, China\\[0pt]}
\end{center}
\vspace{2mm}

\begin{abstract}

\end{abstract}
\vspace{5mm}
We obtain a newly charged BBMB (Bocharova-Bronnikov-Melnikov-Bekenstein) black hole solution from the Einstein-Weyl-Maxwell-conformal (conformally coupled) scalar theory with a positive Weyl coupling parameter $m^2_2$ numerically.
This solution has a primary scalar hair, compared to a secondary scalar hair in the charged BBMB  black hole solution.
The limiting case of  $m_2^2\to\infty$ leads to  the series form of the  charged BBMB black hole solution.
However, for  $m^2_2<0$, we could not obtain any asymptotically flat black hole solution with scalar hair.

\vspace{1.5cm}

\hspace{11.5cm}{Typeset Using \LaTeX}
\newpage
\renewcommand{\thefootnote}{\arabic{footnote}}
\setcounter{footnote}{0}

\section{Introduction}
An extension of Einstein gravity including $R^2$ and $C^2=C_{\mu\nu\rho\sigma}C^{\mu\nu\rho\sigma}$ was renormalizable in Minkowski spacetimes~\cite{Stelle:1976gc}, but it is still suffering from the loss of unitarity due to a massive spin-2 ghost with $m^2_2>0$.  One usually excludes the case of $m^2_2<0$ because it  corresponds to the tachyonic mass.
Recently, the Einstein-Weyl gravity ($R-C^2/2m^2_2$) without having $R^2$ has attracted considerable interest  in obtaining  asymptotically flat black hole solutions numerically.
It was known that the condition of vanishing Ricci scalar has simplified the search for  any static black hole solutions of a Ricci  quadratic gravity.
As a famous example,  a non-Schwarzschild black hole solution found numerically in~\cite{Lu:2015cqa} has a vanishing Ricci scalar and non-vanishing Ricci tensor.
A continued-fraction expansion was used to construct an analytical approximation of numerical black hole solutions found in the Einstein-Weyl gravity~\cite{Kokkotas:2017zwt}.

In this direction, it is meaningful to note that the instability of Schwarzschild black hole (vanishing  Ricci tensor)  is closely related to the appearance of non-Schwarzschild
black hole with non-vanishing Ricci tensor~\cite{Lu:2017kzi,Stelle:2017bdu,Myung:2013doa,Whitt:1985ki}. Explicitly, the small Schwarzschild black hole is unstable against the $s$-mode of Ricci-tensor perturbation (equivalently, massive spin-2 mode)  under the condition of $0<m_2\le m_{\rm th}=0.876/r_+$  with $m_{\rm th}$ and $r_+$ representing the threshold mass for instability and the black hole horizon radius. For  $0<m_2\le m_{\rm th}$, one can find any non-Schwarzschild black hole solution, implying a single branch of non-Schwarzschild black holes.  However, the general problem of connecting the strong field regime to the weak one is not completely understood because  a shooting method was used to determine the weak field region. In other words, one might  successively integrate out further and further approaching asymptotically flat space by tuning the initial conditions near the horizon, resulting in the lack for a systematically asymptotic expansion to the numerical solution~\cite{Goldstein:2017rxn}.   More recently, a multiple shooting method was introduced to handle this problem~\cite{Bonanno:2019rsq}.

On the other hand, considering the Einstein-Maxwell-conformally coupled scalar (EMCS) theory has admitted the charged BBMB black hole whose scalar hair is secondary~\cite{Bekenstein:1974sf}.
This analytic solution might be  regarded as the second counterexample to the no-hair theorem on black holes  even though the scalar hair blows up on the horizon.
Incorporating the Weyl-squared term ($C^2$) into the Einstein-conformally coupled scalar theory (EWCS theory)~\cite{Myung:2019adj}, one could find asymptotic expansions as well as a non-BBMB black hole solution with primary scalar hair. We would like to stress  that  the conformally coupled scalar played an important role in determining asymptotic forms of metric functions.

In this work, we wish to derive a newly charged BBMB black hole solution with primary scalar hair numerically by considering the Einstein-Weyl-Maxwell-conformally coupled scalar (EWMCS) action. If one excludes  the Einstein-Hilbert term, this WMCS theory corresponds to a conformally invariant scalar-vector-tensor theory.  However,   the Einstein-Hilbert term  will be necessary to have a constraint of vanishing Ricci scalar which  plays a crucial role in obtaining numerical black hole  solutions because the EWMCS theory belongs to  a fourth-order gravity theory.

Our paper is organized as follows. We  introduce the EWMCS theory  which is not a conformally invariant scalar-vector-tensor theory in section 2. Section 3 focuses  on obtaining a newly charged BBMB black holes numerically by investigating the case of $m^2_2>0$.   Finally,  we discuss our main  results  in section 4.

\section{EWMCS theory}
We start with the Einstein-Weyl-Maxwell-conformally coupled scalar (EWMCS) action given by
\begin{eqnarray}S_{\rm EWMCS}=\frac{1}{16 \pi G}\int d^4 x\sqrt{-g}
\Big[R-\frac{1}{2m_2^2}C^2-F_{\mu\nu}F^{\mu\nu}-\beta\Big(\phi^2R+6\partial_\mu\phi\partial^\mu\phi\Big)\Big]
\label{EWMCS}
\end{eqnarray}
The second term denotes the Weyl-squared term, called the conformal gravity solely. The conformal coupling parameter $\beta$ is chosen to be $\beta=1/3$ for simplicity.
In limit of $m^2_2\to \infty$, the above action reduces to the Einstein-Maxwell-conformally coupled scalar (EMCS) theory which admitted  a charged BBMB black hole.
Furthermore, excluding the Einstein-Hilbert term from (\ref{EWMCS}) leads to the  WMCS action that is invariant under the conformal transformation
\begin{equation}
g_{\mu\nu} \to \Omega^2(x),~~ \phi \to \frac{\phi}{\Omega},
\end{equation}
where $\Omega(x)$ is an arbitrary positive smooth function of $x$. This implies that adding the Einstein-Hilbert term to the WMCS action breaks the conformal symmetry.
On later, this will gives us an important condition of $R=0$ for obtaining a numerical black hole solution.

The Einstein equation could be derived from (\ref{EWMCS}) as
\begin{equation} \label{nequa1}
E_{\mu\nu}\equiv G_{\mu\nu}-\frac{2}{m_2^2}B_{\mu\nu}-2 T^{\rm M}_{\mu\nu}-T^{\rm \phi}_{\mu\nu}=0,
\end{equation}
where the Einstein tensor  is given by $G_{\mu\nu}=R_{\mu\nu}-Rg_{\mu\nu}/2$ and
the Bach tensor $B_{\mu\nu}$ describes fourth-order terms
\begin{eqnarray}
B_{\mu\nu}=&&R_{\mu\rho\nu\sigma}R^{\rho\sigma}-\frac{1}{4}R^{\rho\sigma}R_{\rho\sigma}g_{\mu\nu}
-\frac{1}{3}R\left(R_{\mu\nu}-\frac{1}{4}R g_{\mu\nu}\right)\nonumber\\
&&+\frac{1}{2}\left(\nabla^2R_{\mu\nu}-\frac{1}{6}\nabla^2R g_{\mu\nu}-\frac{1}{3}\nabla_{\mu}\nabla_{\nu}R\right).
\end{eqnarray}
We note here  that its trace is zero $(B^\mu~_{\mu}=0)$.
The energy-momentum tensors for Maxwell theory and  conformally coupled scalar theory  are  defined by, respectively
\begin{eqnarray} \label{equa2}
T^{\rm M}_{\mu\nu}&=&F_{\mu\rho}F_{\nu}~^\rho- \frac{F^2}{4}g_{\mu\nu},\label{trace} \\
T^{\rm \phi}_{\mu\nu}&=&\frac{1}{3}\Big[\phi^2G_{\mu\nu}+g_{\mu\nu}\nabla^2(\phi^2)-\nabla_\mu\nabla_\nu(\phi^2)
+6\partial_\mu\phi\partial_\nu\phi-3(\partial\phi)^2g_{\mu\nu}\Big].\nonumber
\end{eqnarray}
We note  that $T^{{\rm M} \mu}_\mu=0$ is traceless, but $T^{{\rm \phi} \mu}_\mu=\phi(-\phi R +6\nabla^2\phi)\not=0$.
Also, the Maxwell equation is given by
\begin{equation} \label{maxwell-eq}
\nabla^\mu F_{\mu\nu}=0.
\end{equation}
Apparently, the scalar equation seems to be  a conformally coupled scalar equation
\begin{equation} \label{scalar-eq}
\nabla^2\phi-\frac{1}{6}R\phi=0.
\end{equation}
It is worth noting that  $T^{{\rm \phi} \mu}_\mu=0$ when using (\ref{scalar-eq}), implying the on-shell traceless condition. 
In this case, taking the trace of (\ref{nequa1})   leads to the vanishing Ricci scalar
\begin{equation}
R=0, \label{ricci0}
\end{equation}
which will play an important role in reducing the third-order Einstein equation to the second-order equation.
Taking into account (\ref{ricci0}), one finds a minimally coupled scalar equation
\begin{equation} \label{nscalar-eq}
\nabla^2\phi=0,
\end{equation}
which is found due to the presence of  Einstein-Hilbert term ($R$), breaking the conformal symmetry.
Considering (\ref{nscalar-eq}), it is important to note that the scalar hair has nothing to do with the spontaneous scalarization where acquiring of a scalar hair by the black hole arises from the non-minimal coupling of a scalar to  the Gauss-Bonnet term [$f(\phi){\cal G}$]~~\cite{Doneva:2017bvd,Silva:2017uqg,Antoniou:2017acq} and the Maxwell term [$f(\phi) F^2$]~\cite{Herdeiro:2018wub}.
Imposing  (\ref{ricci0}) and (\ref{nscalar-eq}) on (\ref{nequa1}) leads to the reduced Einstein equation
\begin{eqnarray}
R_{\mu\nu}&=& \frac{2}{m^2_2}\Big(R_{\mu\rho\nu\sigma}R^{\rho\sigma}-\frac{1}{4}R^{\rho\sigma}R_{\rho\sigma}g_{\mu\nu}+\frac{1}{2}\nabla^2R_{\mu\nu}\Big) \nonumber \\
&+& 2T^{\rm M}_{\mu\nu}+\frac{1}{3} [\phi^2R_{\mu\nu}-(\partial \phi)^2g_{\mu\nu} +4\partial_\mu \phi\partial_\nu \phi -2\phi \nabla_\mu\nabla_\nu \phi ].\label{new-eqn}
\end{eqnarray}

Finally, a conformally invariant scalar-vector-tensor (WMCS) theory predicts equations of motion
\begin{equation}
\frac{2}{m_2^2}B_{\mu\nu}=2 T^{\rm M}_{\mu\nu}+T^{\rm \phi}_{\mu\nu},~~\nabla^\mu F_{\mu\nu}=0,~~\nabla^2\phi-\frac{1}{6}R \phi=0,
\end{equation}
which might not be  suitable for admitting a numerical black hole solution with scalar hair because of $ R\not=0$.

\section{Scalar hairy black holes}

\subsection{Charged BBMB black hole solution}

In  limit of $m_2^2\to \infty$, the  action (\ref{EWMCS}) reduces to the one (EMCS theory) admitting  the charged BBMB black hole solution~\cite{Bekenstein:1974sf}
\begin{eqnarray}
&&ds^2_{\rm cBBMP}=-\bar{N}(r)e^{-2\bar{\delta}(r)}dt^2+\frac{dr^2}{\bar{N}(r)}+r^2d\Omega^2_2, \nonumber \\
&&\bar{N}(r)=\Big(1-\frac{m}{ r}\Big)^2,~~\bar{\delta}(r)=0,~~\bar{ \phi}(r)=\frac{\sqrt{3(m^2-Q^2)}}{r-m},~~\bar{A}_t=\bar{v}(r)=\frac{Q}{r}-\frac{Q}{r_+}, \label{bbmb1}
\end{eqnarray}
where $m$ is the mass of the black hole and $Q$ denotes the Maxwell charge. This line element takes  the same form as in the extremal Reissner-Nordstr\"om (RN) black hole, but
the scalar hair blows up at the horizon $r=m$. Also, we note that  $\bar{\phi}(r)$ belongs to the secondary hair because it does not have an independent scalar charge $Q_s$.
It was shown   forty years ago that this black hole is unstable against the scalar perturbation since it belongs to an extremal RN black hole~\cite{Bronnikov:1978mx}.
We would like to stress that the charged BBMB solution (\ref{bbmb1}) was obtained under the conditions of vanishing Ricci scalar and non-vanishing Ricci tensor, in addition to non-vanishing scalar. Also, the extremality of  (\ref{bbmb1}) reflects the conformal scalar coupling to the Einstein-Hilbert term~\cite{Xanthopoulos:1991mx,Tomikawa:2017vun}.

On the other hand, it is worth noting that  a non-extremal black hole with constant scalar hair could be found  from the EMCS theory~\cite{Astorino:2013sfa}.

\subsection{Newly charged BBMB black holes}
Now, we wish to derive a newly charged BBMB black hole solution which will carry a primary scalar hair by including the Weyl-squared term.
We remind the reader that (\ref{bbmb1}) is an analytic solution found from the EMCS theory. It is legitimate to say that one could not derive any analytic solution from the EWMCS theory because of the Weyl-squared term. Hence, we would be better to look for a numerical black hole solutions in the EWMCS theory.
In this case, we split  $m_2^2$ into three cases: $m_2\to \infty$, $m^2_2>0$, and $m^2_2<0$.

In order to find scalarized charged black holes numerically, one proposes the metric ansatz
\begin{eqnarray}
&&ds^2_{\rm scbh}=-N(r)e^{-2\delta(r)}dt^2+\frac{dr^2}{N(r)}+r^2d\Omega^2_2,~~\phi(r)\not=0,~~A_t=v(r), \label{sRN}
\end{eqnarray}
where $\delta(r)\not=0$ will describe  a newly charged  BBMB black hole.
Substituting Eq.(\ref{sRN}) into Eqs.(\ref{nequa1}), (\ref{maxwell-eq}),
and (\ref{nscalar-eq}), one finds four equations for $N(r),~\delta(r),~\phi(r),$ and $v(r)$ as 
\begin{eqnarray}
&&N''r^2[rN'-2N(1+r\delta')]+N'r[rN'(4+r\delta')-2]
-2N[2+rN'(4r\delta'+2r^2\delta'^2-1)]\nonumber\\
&&+N^2(4+6r^2\delta'^2+4r^3\delta'^3)+\frac{2m_2^2r^2}{3}[(\phi^2(rN'-1)-3(rN'+e^{2\delta}r^2V'^2-1)\nonumber\\
&&+r^2N'\phi\phi'+N(6r\delta'-3+\phi^2(1-2r\delta')-2r\phi(r\delta'-2)\phi'+3r^2\phi'^2)]=0,\label{neweq-1}\\
&&2+rN'(3r\delta'-4)-r^2N''+N(2r^2\delta''-2r^2\delta'^2+4r\delta'-2)=0,\label{neweq-2}\\
&&rN'\phi'+N[(2-r\delta')\phi'+r\phi'']=0,\label{neweq-3}\\
&&Q+e^{\delta}r^2v'=0. \label{neweq-3}
\end{eqnarray}
In deriving (\ref{neweq-1}), we wish to mention the Einstein equation (\ref{nequa1})  that $E_{tt}=0$: fourth-order equation; $E_{rr}=0$: third-order equation;  $E_{\theta\theta}=0$: fourth-order equation. We obtain one third-order equation $N'''(r)+\cdots=0$ by making use of two fourth-order equations.
Considering  $E_{rr}=0$ together  with $N'''(r)+\cdots=0$ leads to the second-order equation (\ref{neweq-1}). During this process,   we use  $R=0$  (\ref{neweq-2}) to eliminate $\delta''(r)$ in (\ref{neweq-1}).
At this stage, we have to mention that one arrives at the same equation as (\ref{neweq-1}) when using the reduced Einstein equation (\ref{new-eqn}).

First of all, we introduce  the near-horizon expansion  by considering the charged BBMB solution (\ref{bbmb1})
\begin{eqnarray}\label{CBBMBexpr2}
N(r)&=&\sum^{\infty}_{i=2} N_i(r-r_+)^i,\quad
\delta(r)=\sum^{\infty}_{i=0}\delta_i(r-r_+)^i,\nonumber\\
v(r)&=&\sum^{\infty}_{i=1}v_i(r-r_+)^i,\quad
\psi(r)=\sum^{\infty}_{i=1}\psi_i(r-r_+)^i,
\end{eqnarray}
where  `$i=2$' in $N(r)$ reflects the conformally  coupled scalar background solution.
Also it is worth noting that $\psi=1/\phi$ is introduced in the near-horizon for technical reason.
The first four coefficients are computed as
\begin{eqnarray}
N_2&&=\frac{1}{r_+^2},~~ N_3=-\frac{6(r_+^2-Q^2)\psi_1^2+4}{3r_+^3},~~N_4=\frac{5}{3r_+^4}
+\frac{5(r_+^2-Q^2)\psi_1^2}{r_+^4}-\frac{3(r_+^2-Q^2)^2\psi_1^4}{r_+^4},\nonumber\\
N_5&&=-\frac{86}{45r_+^5}-\frac{(37r_+^2-46Q^2)\psi_1^2}{5r_+^5}-\frac{27(r_+^2-Q^2)(2Q^2-r_+^2-\frac{6}{m_2^2})\psi_1^4}{5r_+^5}
-\frac{6(r_+^2-Q^2)^3\psi_1^6}{r_+^5},\nonumber\\
\underline{\delta_0},&&  \delta_1=\frac{2+6(Q^2-r_+^2)\psi_1^2}{3r_+},~~
\delta_2=-\frac{1+15(Q^2-r_+^2)\psi_1^2+36(Q^2-r_+^2)^2\psi_1^4}{9r_+^2},\nonumber\\
\delta_3&&=\frac{34}{405r_+^3}+\frac{2(38Q^2-11r_+^2)\psi_1^2}{45r_+^3}
+\frac{2(Q^2-r_+^2)(59Q^2-32r_+^2-\frac{162}{m_2^2})\psi_1^4}{15r_+^3}
+\frac{32(Q^2-r_+^2)^3\psi_1^6}{3r_+^3},\nonumber\\
v_1&&=-\frac{e^{-\delta_0}Q}{r_+^2},~~
v_2=\frac{e^{-\delta_0}Q}{r_+^3}\Big[\frac{4}{3}-(r_+^2-Q^2)\psi_1^2\Big],\label{ncoef2}\\
v_3&&=-\frac{e^{-\delta_0}Q}{r_+^4}\Big[\frac{14}{9}-\frac{7}{3}(r_+^2-Q^2)\psi_1^2+2(r_+^2-Q^2)^2\psi_1^4\Big],\nonumber\\
v_4&&=\frac{e^{-\delta_0}Q}{r_+^5}\Big[\frac{232}{135}-\frac{(107r_+^2-116Q^2)\psi_1^2}{30}
+\frac{(r_+^2-Q^2)(59r_+^2-68Q^2+\frac{54}{m_2^2})\psi_1^4}{10} \nonumber\\
&&\quad+5(r_+^2-Q^2)^3\psi_1^6\Big],\nonumber\\
\underline{\psi_1},&& \psi_2=-\frac{\psi_1[1-3(r_+^2-Q^2)\psi_1^2]}{3r_+},~~ \psi_3=\frac{2\psi_1[1-3(r_+^2-Q^2)\psi_1^2]^2}{9r_+^2},\nonumber\\
\psi_4&&=-\frac{22\psi_1}{135r_+^3}+\frac{(37r_+^2-46Q^2)\psi_1^3}{30r_+^3}-\frac{3(r_+^2-Q^2)(13r_+^2-16Q^2+\frac{18}{m_2^2})\psi_1^5}{10r_+^3}
-\frac{5(r_+^2-Q^2)^3\psi_1^7}{r_+^3},\nonumber
\end{eqnarray}
where $Q$ denotes the $U(1)$ charge.
We note that $\{N_5,~\delta_3,~v_4,\psi_4\}$ contain the Weyl-squared term with $1/m^2_2$.
In our theory, two free parameters $\delta_0$ and $\psi_1$ will be determined
when matching (\ref{CBBMBexpr2}) with the following asymptotic expansions
for $r\gg r_+$:
\begin{eqnarray}\label{insol2}
N(r)&=&1-\frac{2m}{r}+\frac{Q^2+Q_s^2}{r^2}+\frac{2}{3r^4}\Big[(Q^2+Q_s^2)\Big(Q_s^2+\frac{6}{m_2^2}\Big)-m^2Q_s^2\Big]+\cdots, \nonumber\\
\delta(r)&=&\frac{1}{6r^4}\Big[(Q_s^2+Q^2)\Big(Q_s^2+\frac{6}{m_2^2}\Big)-m^2Q_s^2\Big]-\frac{4mQ_s^2
(m^2-Q^2-Q_s^2)}{5r^5}+\cdots,\nonumber\\
v(r)&=&-\frac{Q}{r_+}+\frac{Q}{r}-\frac{Q}{30r^5}\Big[(Q_s^2+Q^2)\Big(Q_s^2+\frac{6}{m_2^2}\Big)-m^2Q_s^2\Big]+\cdots, \nonumber\\
\phi(r)&=&\frac{\sqrt{3}Q_s}{r}+\frac{\sqrt{3}Q_s m}{r^2}+\frac{\sqrt{3}Q_s}{3r^3}(4m^2-Q^2-Q_s^2)
+\frac{\sqrt{3}Q_sm}{r^4}(2m^2-Q^2-Q_s^2)\nonumber\\
&+&\frac{\sqrt{3}Q_s}{10r^5}\Big[32m^4-m^2(24Q^2+23Q_s^2)+(Q^2+Q_s^2)\Big(2Q^2+Q_s^2-\frac{6}{m_2^2}\Big)\Big]+\cdots,
\end{eqnarray}
where $m$ is the ADM (Arnowitt-Deser-Misner) mass and $Q_s$  represents the scalar charge.
Here, we observe that ${\cal O}(r^{-3})$ is missed in $N(r),~\delta(r),$ and $v(r)$, implying  the feature of a conformally coupled scalar in the EWMCS theory.
In addition, we note  that asymptotic expansions are not available if the Maxwell and scalar fields are turned off.
This  explains clearly  why the asymptotic expansion is not available in the Einstein-Weyl gravity~\cite{Lu:2015cqa}.

At this stage, it is important to check whether the limit of $m^2_2\to \infty$ recovers the charged BBMB black hole solution (\ref{bbmb1}).
We stress that (\ref{bbmb1}) is an analytic solution, whereas our way of obtaining solution is a numerical approach.
So, we wish to obtain a numerical charged BBMB black hole solution.
In this case, we obtain three relations by examining numerical computation~\cite{Zou:2019ays} as
\begin{equation}
r_+\approx m,~~m \approx \sqrt{Q^2+Q_s^2},~~\psi_1\approx \frac{1}{\sqrt{3(r_+^2-Q^2)}},~~\delta_0=0.0001.  \label{relation}
\end{equation}
Plugging (\ref{relation}) into (\ref{ncoef2}), we recover the near-horizon expansion for the charged BBMB solution (\ref{bbmb1})
\begin{eqnarray}\label{ncoef3}
&&\bar{N}(r)=\frac{(r-r_+)^2}{r_+^2}-\frac{2(r-r_+)^3}{r_+^3}+\frac{3(r-r_+)^4}{r_+^4}+... \to \Big[\Big(1-\frac{r_+}{r}\Big)^2\Big]_{r=r_+},\nonumber\\
&&\bar{v}(r)=-\frac{Q(r-r_+)}{r_+^2}+\frac{Q(r-r_+)^2}{r_+^3}-\frac{Q(r-r_+)^3}{r_+^4}+...\to \Big[\frac{Q}{r}\Big]_{r=r_+}-\frac{Q}{r_+},\nonumber\\
&&\bar{\delta}(r)=0,\nonumber \\
&&\bar{\psi}(r)=\frac{r-r_+}{\sqrt{3(r_+^2-Q^2)}}.
\end{eqnarray}
On the other hand, substituting  (\ref{relation}) into (\ref{insol2})  leads to  the asymptotic expansion for the charged BBMB solution (\ref{bbmb1})
\begin{eqnarray}\label{insol3}
&&\bar{N}(r)=1-\frac{2m}{r}+\frac{m^2}{r^2}=\Big(1-\frac{m}{r}\Big)^2,  \nonumber \\
&&\bar{v}(r)=\frac{Q}{r}-\frac{Q}{r_+},  \nonumber \\
&&\bar{\delta}(r)= 0\nonumber \\
&&\bar{ \phi}(r)=\sqrt{3(m^2-Q^2)}\Big[\frac{1}{r}+\frac{m}{r^2}+\frac{m^2}{r^3}+\frac{m^3}{r^4}+\frac{m^4}{r^5}\cdots\Big] \to \Big[\frac{\sqrt{3(m^2-Q^2)}}{r-m}\Big]_{r\gg m}.
\end{eqnarray}
Here, we observe that (\ref{insol3}) does not contain a scalar charge $Q_s$ explicitly. This  implies that the scalar hair is secondary in the charged BBMB solution.
Two expressions (\ref{ncoef3}) and (\ref{insol3}) constitute   a numerical charged BBMB black hole solution found from the EMCS theory.

Now, let us consider the $m^2_2>0$ case.
Explicitly, we  may choose the horizon radius $r_+=0.75$, $U(1)$ charge $Q=0.5$ together with $m_2^2=2$  to construct a newly charged  BBMB  black hole with the ADM mass $m=0.7974$  and scalar charge $Q_s=0.5766$.  This implies a relation  of $m^2\not\approx Q^2+Q_s^2$, indicating a non-extremal black hole.
As are shown in Figs. 1 and 2, the metric functions, scalar hair, and vector potential are compared  to those for the charged BBMB black hole without Weyl-squared term.
The appearance of non-zero $\delta(r)$ and scalar charge $Q_s$  indicates  clearly that this charged BBMB solution has a primary scalar hair.
This represents  a newly charged BBMB black hole which is quite different from the charged BBMB solution in the single branch of $m^2_2>0$.
\begin{figure*}[t!]
   \centering
   \includegraphics{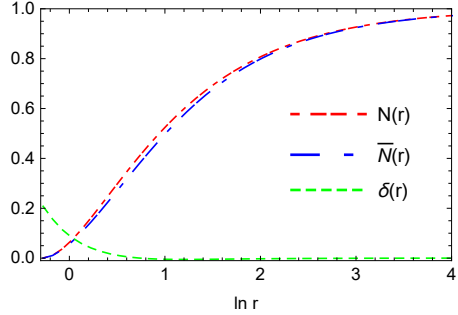}
\caption{Plot of metric function $N(r)$ for a newly charged BBMB solution with $m_2^2=2$, compared to $\bar{N}(r)$ for the charged BBMB solution with $r_+=m=0.75$ and $ Q_s=0.5590$. Here, the horizon radius is located at $r_+=0.75$ (implying $\ln r_+=-0.274$)
together with charge $Q=0.5$.  $\delta(r)$ starts with  $\delta_0=0.215$, while it is always zero [$\bar{\delta}(r)=0$] for the charged BBMB solution. }
\end{figure*}
\begin{figure*}[t!]
   \centering
     \includegraphics{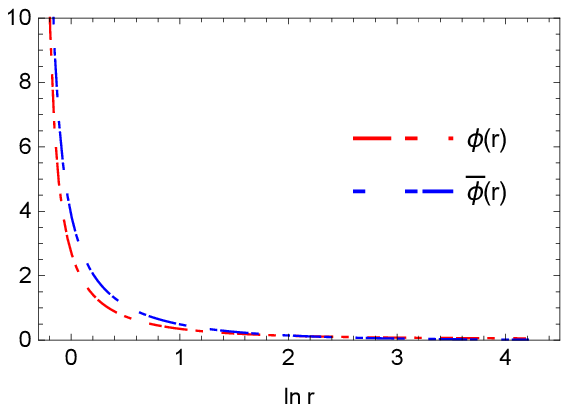}
     \hfill%
     \includegraphics{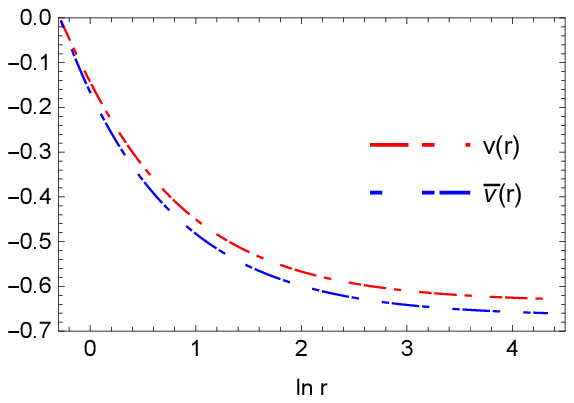}
    \hfill%
\caption{(Left) Plot of scalar hair  $\phi(r)$ for a newly charged BBMB solution with $\psi_1=1.3803$ compared to $\bar{\phi}(r)$ for the charged BBMB solution with $\psi_1=1.0327$.
(Right) Plot of vector potential $A_t=v(r)$ for a newly charged BBMB solution, comparing with $\bar{A}_t(r)=\bar{v}(r)$ for the charged BBMB solution.}
\end{figure*}

However, for $m^2_2<0$, we employ $m^2_2=-1.5$, $\psi_1=0.7183$, and $\delta_0=-0.4$ to derive a numerical solution $\{N_n(r),~\delta_n(r),~v_n(r),~\phi_n(r)\}$ shown in Fig. 3.
We stress  that one could not obtain any asymptotically flat numerical black hole  solutions for $m^2_2<0$ because $N(r)$ diverges  as $r$ increases.
In Table 1, we list the several cases that  are used for finding divergent metric functions $N(r)$. Actually, a numerical set of $\{m^2_2,~\psi_1,~\delta_0\}$ determines  the near-horizon expansion.
 Even though these do not represent a complete set  for the case of $m^2_2<0$,
it is reasonable to say that  any asymptotically flat numerical black hole  solutions are not allowed for $m^2_2<0$  because of $m^2_2\in [-0.0008,-1500]$.

\begin{figure*}[t!]
   \centering
   \includegraphics{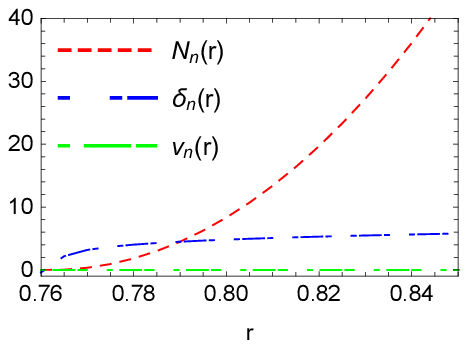}
     \hfill%
   \includegraphics{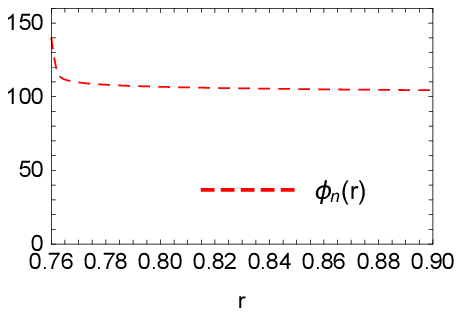}
\caption{(Left) Plot of divergent metric function $N_n(r)$ for a charged BBMB solution with $m_2^2=-1.5$ and  $\psi_1=0.7183$. Here, the horizon radius is located at $r_+=0.75$ together with $U(1)$ charge $Q=0.5$.  $\delta_n(r)$ starts with  $\delta_0=-0.4$. (Right) Plot for scalar hair $\phi_n(r)$ with divergent behavior near the horizon.  }
\end{figure*}

\begin{table}[h]
\begin{tabular}{|c|c|c|}
  \hline
  $m_2^2$ & $\psi_1$ & $\delta_0$ \\ \hline
  $-0.0008$ &0.1017 & $-1/110$ \\ \hline
  $-0.0025$ &0.1366 & $-1/110$ \\ \hline
  $-0.0054$ &0.1438 & $-1/110$ \\ \hline
  $-0.0217$ &0.1825 & $-1/110$ \\ \hline
  $-0.0357$ &0.2676 & $-1/88$ \\ \hline
  $-0.0555$ &0.2712 & $-1/60$ \\ \hline
  $-0.0833$ &0.3207& $-1/38$ \\ \hline
  $-0.125$ &0.3423& $-1/28$\\ \hline
  $-0.2$ &0.4027& $-1/18$\\ \hline
  $-0.75$ &0.6254& $-1/4$\\ \hline
  $-2$ &0.7686& $-1/2$\\ \hline
  $-12$ &0.8838& $ -1$\\ \hline
  $-32$ &0.9854& $-2$\\ \hline
  $-92$ &1.3089& $-4$\\ \hline
  $-150$ &1.4750& $-5$\\ \hline
  $-1500$ &2.0125& $-7$\\ \hline
\end{tabular}
\caption{List used for finding divergent metric functions $N(r)$ for $m^2_2<0$. }
\end{table} 
\section{Discussions}

A minimally coupled scalar does not obey the Gauss-law and thus, a black hole
cannot have a scalar hair~\cite{Herdeiro:2015waa}. This describes no-hair theorem on black holes.
 The scalar-tensor theories admitting other than the Einstein gravity are rather rare.
 However, it is known that  introducing  the Einstein-conformally coupled scalar (ECS) theory lead to the BBMB black hole with scalar hair
although it is not a conformally invariant theory~\cite{Bocharova:1970skc,Bekenstein:1974sf}. This BBMB scalar hair is secondary and blows up on the horizon.
The BBMB  solution corresponds to  the first counterexample to the no-hair theorem on black holes. Also, we note that the  BBMB solution with  scalar hair in the ECS theory does not have a continuous limit to the Schwarzschild black hole without scalar hair found in the Einstein theory.
In this case, it is surprised that the scalar equation (\ref{scalar-eq}) becomes (\ref{nscalar-eq}) when imposing the condition of $R=0$.
In  case of $R\not=0$ (without the Einstein-Hilbert term $R$), no-hair theorem is still  valid for a massive scalar  with
non-minimal coupling to gravity~\cite{Hod:2017ssh}. We point out  a difference that the original scalar equation $\nabla^2 \phi=0$ and the final scalar equation $\nabla^2 \phi=0$ obtained after using $R=0$. Even though these are the same minimally coupled equation, the former could not implement black hole with scalar hair whereas the latter admits black hole with scalar hair.  It is important to note that the BBMB scalar hair has nothing to do with the spontaneous scalarization where acquiring of a smooth scalar hair by the black hole arises from the non-minimal coupling of a scalar to  the Gauss-Bonnet term [$f(\phi){\cal G}$]~~\cite{Doneva:2017bvd,Silva:2017uqg,Antoniou:2017acq} and the Maxwell term [$f(\phi) F^2$]~\cite{Herdeiro:2018wub}.

On the other hand, it is well-known  that the vanishing Ricci scalar ($R=0$) played an important role in obtaining numerical non-Schwarzschild black hole solution in the Einstein-Weyl gravity~\cite{Lu:2015cqa} which  includes a conformally invariant Weyl tensor and a traceless Bach tensor in equation. Such a construction of numerical black hole could be extended to include a Maxwell kinetic term because its energy-momentum tensor is traceless~\cite{Lin:2016kip}.

Furthermore, we would like to mention that  the charged BBMB solution (\ref{bbmb1}) with scalar hair obtained in the EMCS theory does not have a
continuous limit to the RN black hole found in the Einstein-Maxwell theory, showing a feature of conformally coupled
scalar to gravity~\cite{Bekenstein:1974sf}.  This might be  considered  as the second counterexample to the no-hair theorem on black holes  even though the scalar hair blows up on the horizon.

In this work,  we have shown that adding the Weyl-squared term ($C^2$) with parameter $1/m^2_2$ to the EMCS theory (EWMCS theory) may admit
a charged BBMB solution with three of ADM mass $m$, Maxwell charge $Q$, and scalar charge $Q_s$.
For $m^2_2>0$, we have obtained a newly charged BBMB black hole solution  numerically from the EWMCS theory  which is not a conformally invariant scalar-vector-tensor theory.
This solution includes  a primary scalar hair, compared to a secondary scalar hair in the charged BBMB  black hole solution found from the EMCS theory.
We emphasize that the Weyl-squared term  makes
a shift from the charged BBMB solution with secondary scalar hair to the newly charged BBMB solution with primary scalar hair.
Interestingly, the limit of   $m_2^2\to\infty$ leads to  the series solution of the  charged BBMB black hole.
However, for  $m^2_2<0$, we could not obtain any asymptotically flat black hole solution with scalar hair.
This may be  so because the case of $m^2_2<0$ corresponds to tachyonic mass for a massive spin-2 mode when perturbing around the Minkowski background.

Lastly, we mention that  a recent work~\cite{Caceres:2020myr} has explored  similar problems with different techniques.

\vspace{1cm}
{\bf Acknowledgments}

 \vspace{1cm}

This work was supported by the National Research Foundation of Korea (NRF) grant funded by the Korea government (MOE)
(No. NRF-2017R1A2B4002057).

\end{document}